\documentclass{elsart}
\usepackage{graphicx,amsmath,amsfonts,amssymb,bbm,psfrag,url}
\usepackage[T1]{fontenc}
\newcommand{\dd}{\mathrm{d}}
\begin {document}
\begin{frontmatter}
\title{Stationary state of a heated granular gas:  
fate of the usual $H$-functional}
\author[ioana]{Ioana Bena},
\author[ioana]{Fran\c{c}ois Coppex},
\author[ioana]{Michel Droz},
\author[emmanuel,paolo]{Paolo Visco},
\author[emmanuel]{Emmanuel Trizac},
\author[frederic]{Fr\'ed\'eric van Wijland}
\address[ioana]{Department of Theoretical Physics, 
University of Geneva, CH-1211
Geneva 4, Switzerland}
\address[emmanuel]{Laboratoire de Physique Th\'eorique et
  Mod\`eles Statistiques, UMR CNRS 8626, B\^atiment 100, Universit\'e
  Paris-Sud, 91405 Orsay Cedex, France}
\address[paolo]{Laboratoire de Physique Th\'eorique, UMR CNRS 8627, B\^atiment 210,
  Universit\'e Paris-Sud, 91405 Orsay Cedex, France}
\address[frederic]{Laboratoire Mati\`ere et Syst\`emes Complexes, CNRS UMR 7057,
  Universit\'e Denis Diderot, 2 place Jussieu, 75251 Paris Cedex 05, France}

\begin{abstract}
  We consider the characterization of the nonequilibrium stationary state of a
  randomly-driven granular gas in terms of an entropy-production based
  variational formulation. Enforcing spatial homogeneity, we first consider
  the temporal stability of the stationary state reached after a transient. In
  connection, two heuristic albeit physically motivated candidates for the
  non-equilibrium entropy production are put forward. It turns out that none of them
  displays an extremum for the stationary velocity distribution selected by
  the dynamics. Finally, the relevance of the relative Kullbach entropy is 
  discussed.
\end{abstract}
\end{frontmatter}

\section{Introduction}
\label{introduction}

Apart from being the subject of intense experimental activity, granular gases
are also a particularly fertile testing ground for new theoretical ideas and
problems, especially within the field of nonequilibrium statistical physics.
One such a problem is the role of entropy production as a Lyapunov functional
for nonequilibrium steady-states.  This problem has its roots in the fifties,
in the works of the Brussels group around Prigogine~\cite{prigogine} on the
minimum entropy production theorem. The limitations of this theorem,
that relies essentially on the linear response formalism (i.e., has a domain of validity
that is restricted to close-to-equilibrium situations), were rather clear
already at that time, and a first extension to far-from-equilibrium situations
was proposed under the form of  the phenomenological ``general evolution criterion" of Glansdorff and Prigogine
(see~\cite{glansdorff} and references therein).

Since then, there was steady effort, and a huge body of literature, in the search
for a variational principle for steady-states that are arbitrarily far from
equilibrium. Several candidates for a ``nonequilibrium entropy production"
with extremal properties at stationarity were proposed in different contexts, and at various levels of coarse-graining of the description -- from the microscopic to the phenomenological ones. Some success was encountered for Markovian systems described by a master equation for the probability distribution function of the microstates -- starting with the pioneering work of Jiu-li {\em et al}~\cite{jiu}, and intensively studied afterwards (see, e.g.,~\cite{bertini,gaspard}  to cite only a few).
Also, a connection between the phase space contraction rate in dissipative, externally driven systems and an entropy production rate was established in some cases,
see e.g.~\cite{nicolis} for a critical discussion.
An extension of Jaynes' maximum entropy inference principle
(MaxEnt) to nonequilibrium situations was proposed~\cite{jaynes}, and
illustrated recently on several examples~\cite{maxent}. The resulting picture
is, however, rather confusing and sometimes even contradictory (e.g., some of
the above-mentioned papers speak of a ``maximum" entropy production rate at
stationarity, while others refer to a ``minimum").

One of the main difficulties of nonequilibrium statistical mechanics is the
scarcity of solvable models, on the basis of which one could, eventually, get
some clarification on these controversial points.  The purpose of the present
work is to consider such a solvable model, namely a granular gas modeled as an assembly of inelastic hard-spheres with {\em constant restitution coefficient}, in which energy is injected by means of random forces acting independently upon the particles. 
The balance between dissipation and the random kicks allows the system to reach a nonequilibrium steady-state (NESS). In a Boltzmann equation description,
one can compute (in some perturbative expansion) the single-particle probability distribution function (pdf). 
This model is widely-used and very successful in explaining many features of granular 
systems (see, e.g.,~\cite{noije,MontaneroSantos,BP}). One of the question is thus whether this model  is also appropriate in describing {\em thermodynamical} properties of granular systems -- in particular, the entropy production rate and its eventual relationship with the relaxation to  NESS.
We propose two heuristic -- albeit physically motivated -- candidates for the
nonequilibrium entropy production rate, as functionals of the pdf, and we discuss
their extremal properties in  NESS. Such a granular gas has a
strong ``built-in" irreversible element at the very level of the grain
dynamics, which is represented by the inelasticity of the collisions. However,
one may  ask whether in the limit of a very weak inelasticity 
(i.e., for steady-states that
are arbitrarily ``close to equilibrium") one could recover a kind of ``minimum
entropy production theorem" in a stochastic formulation  -- an equivalent of that described in~\cite{jiu}. 
We will also address this point here.

In the next section we are introducing the model, and in Sec.~3 we study 
the nonequilibrium steady-state and its linear stability. Section~4
is devoted to the discussion of the nonequilibrium entropy production issue,
and the behavior of the relative Kullback entropy.
We conclude in Sec.~5 with a brief discussion of the limitations of this model
as far as describing the thermodynamics.

\section{The model}
\label{model}

We consider a granular gas of inelastic hard spheres 
in dimension $d\geqslant 2$, 
uniformly heated by a stochastic thermostat, 
as described in detail in~\cite{noije,MontaneroSantos}.
The particles undergo binary inelastic collisions, modeled through
a {\em constant restitution coefficient} $\alpha \in [0,1]$ that
is meant to characterize the {\em degree of inelasticity}; the limit
$\alpha=1$ corresponds to perfectly elastic collisions, while $\alpha=0$ 
corresponds to the perfect inelastic ones. 
Each particle $i$ (of mass $m$) is subjected to an external 
Gaussian white noise force $\boldsymbol{\xi}_i(t)$;
these forces are uncorrelated for different particles,
and homogeneous in space,
\begin{equation}
\langle {\xi}_{i,\alpha}(t){\xi}_{j,\beta}(t')\rangle 
=m^2{\xi_0^2}\delta_{ij}\delta_{\alpha \beta}
\delta(t-t')\,,\;\alpha,\beta=1,...,d\,.
\end{equation}
We describe the system at the level of the kinetic theory, and
for simplicity, without affecting the overall conclusions,
we shall concentrate on the {\em spatially homogeneous} case.
For the single particle distribution function 
$f(\mathbf{r},\,\mathbf{v}_1,\,t)=f(\mathbf{v}_1,\,t)$, 
the Boltzmann equation reads then:
\begin{equation}
\partial_t f(\mathbf{v}_1,t) = \chi I[f,f] + 
\frac{\xi_0^2}{2} \frac{\partial^2}{\partial \mathbf{v}_1^2} 
f(\mathbf{v}_1,t)\,.
\label{eq1}
\end{equation}
The extra  term $({\xi_0^2}/{2})
({\partial^2}/{\partial \mathbf{v}_1^2}) f(\mathbf{v}_1,t)$ accounts for the
change in the distribution function caused by the random ``kicks" the external
thermostat is applying on the grains. It corresponds to an injection of
energy at constant rate $d\xi_0^2/2$ per unit mass.  $\chi$ is the pair
correlation function at contact  and
\begin{multline}
I[f,f] = \sigma^{d-1} \int_{\mathbbm{R}^d} \dd \mathbf{v}_2 
\int \dd \widehat{\boldsymbol{\sigma}} \, 
\theta(\widehat{\boldsymbol{\sigma}} \cdot \mathbf{v}_{12}) 
(\widehat{\boldsymbol{\sigma}} \cdot \mathbf{v}_{12}) \\
\times \left( \frac{1}{\alpha^2} b^{-1} -1 \right) 
f(\mathbf{v}_1,t) f(\mathbf{v}_2,t)
\label{eq2}
\end{multline}
represents the inelastic two-particle collision operator. Here $\sigma$ is the
diameter of the hard spheres; $\widehat{\boldsymbol{\sigma}}$ is a unit vector
joining the centers of the particles at contact; $\mathbf{v}_{12} =
\mathbf{v}_1 - \mathbf{v}_2$; $\theta(...)$ is the Heaviside step-function;
and $b^{-1}$ is an operator that restitutes the pre-collisional velocities, i.e.,
\begin{subequations}
\begin{eqnarray}
b^{-1} \mathbf{v}_1 = \mathbf{v}_1^{**} = \mathbf{v}_1 - 
\frac{1+\alpha}{2\alpha} (\mathbf{v}_{12} 
\cdot \widehat{\boldsymbol{\sigma}}) 
\widehat{\boldsymbol{\sigma}}\,, 
\label{resume4a}\\
b^{-1} \mathbf{v}_2 = \mathbf{v}_2^{**} = \mathbf{v}_2 + 
\frac{1+\alpha}{2\alpha} (\mathbf{v}_{12} 
\cdot \widehat{\boldsymbol{\sigma}}) \widehat{\boldsymbol{\sigma}}\,. 
\label{resume4b}
\end{eqnarray}
\label{resume4}
\end{subequations} Note that the post-collisional velocities are
\begin{subequations}
\begin{eqnarray}
b \mathbf{v}_1 = \mathbf{v}_1^{*} = \mathbf{v}_1 - 
\frac{1+\alpha}{2} (\mathbf{v}_{12} \cdot \widehat{\boldsymbol{\sigma}}) 
\widehat{\boldsymbol{\sigma}}\,, 
\label{resume4a2}\\
b \mathbf{v}_2 = \mathbf{v}_2^{*} = \mathbf{v}_2 + \frac{1+\alpha}{2} 
(\mathbf{v}_{12} \cdot \widehat{\boldsymbol{\sigma}}) 
\widehat{\boldsymbol{\sigma}}\,. \label{resume4b2}
\end{eqnarray}\label{resume42}
\end{subequations}

\section{Scaling solution and stationary state}

\subsection{Scaling solution of Boltzmann's equation}

It turns convenient to introduce the pdf $\widetilde f$ of rescaled velocities
$ \mathbf{c}=\mathbf{v}/v_T$:
\begin{equation}
f(\mathbf{v},t) = \frac{n}{v_T(t)^d} \widetilde{f}(c, t), 
\label{solutionechelle}
\end{equation}
where $n$ is the number particle density and
\begin{equation}
v_T(t)=\sqrt{\frac{2 k_B T(t)}{m}}
\end{equation}
is the thermal velocity associated to the kinetic 
temperature of the particles,
\begin{equation}
\frac{d}{2}  k_B T(t) = \frac{1}{n}\,
\int_{\mathbbm{R}^d} \dd \mathbf{v} \frac{m}{2} v^2 
f(\mathbf{v},t)
\label{new1}
\end{equation}
($k_B$ is Boltzmann's constant).

For inelastic collisions,  
$\widetilde{f}(c,t)$ is different from a Gaussian
\begin{equation}
\phi(c)=\frac{1}{\pi^{d/2}} \mathrm{e}^{-c^2} \,,
\end{equation}
and it is customary to characterize
its deviation from a Gaussian
through a series development in terms of Sonine polynomials $S_n(c^2)$,
which, in practice, is truncated to the first non-zero term \cite{BP},
\begin{equation}
\widetilde{f}(c,t) = \phi(c)\,
\left[ 1+ a_2(t) S_2(c^2)\right]\,,
\label{correction}
\end{equation}
where
\begin{equation}
S_2(c^2) = \frac{1}{2} c^4 - \frac{d+2}{2} c^2 + \frac{d(d+2)}{8}.
\end{equation}
The possible explicit temporal dependence of $\widetilde{f}(c,t)$ appears
through the time-dependent coefficient $a_2(t)$ of the Sonine polynomial
$S_2(c^2)$.


For consistency of the description, it is found that the kinetic temperature
$T(t)$ and the coefficient $a_2(t)$ obey a set of two coupled nonlinear
first-order differential equations:
\begin{multline}
\displaystyle\frac{\dd T(t)}{\dd t} = \frac{m \xi_0^2}{k_B} - \sqrt{\frac{k_B}{\pi m}} \frac{n \chi \sigma^{d-1}  
(1-\alpha^2)S_d}{d} \,T^{3/2}(t)\\
\times \left[ 1 + \frac{3}{16} a_2(t) + \frac{9}{1024} a_2^2(t) \right]\,, 
\label{nonlinear1}
\end{multline}
\begin{multline}
  \displaystyle\frac{\dd a_2(t)}{\dd t}+ \frac{2m \xi_0^2}{k_BT(t)}\,a_2(t)+\sqrt{\frac{k_B T(t)}{\pi m}}\,\frac{4n\chi \sigma^{d-1}(1-\alpha^2)S_d}{d(d+2)}\\
  \times \left[ 1 + \frac{3}{16} a_2(t) + \frac{9}{1024} a_2^2(t) \right] \left[1+\frac{d(d+2)}{8}a_2(t)\right] \\
  =\sqrt{\frac{2k_BT(t)}{\pi m}} \,\frac{4n\chi \sigma^{d-1}S_d}{d(d+2)} \,\left[\frac{1-\alpha^2}{1+\alpha^2}+D_1 a_2(t)+D_2 a_2^2(t)\right]\,.\\
\label{nonlinear2}
\end{multline}
Here $S_d= 2 \pi^{d/2} / \Gamma(d/2)$ is the surface of the unit-radius sphere
in dimension $d$, $\Gamma$ being Euler's Gamma function.
Equation~(\ref{nonlinear1}) follows from the definition~(\ref{new1}) of the
kinetic temperature, while Eq.~(\ref{nonlinear2}) is obtained from the limit
of vanishing velocities of the Boltzmann equation~(\ref{eq1}),
see~\cite{limite}.  The coefficients $D_1$ and $D_2$ are given, respectively,
by~\cite{limite}:
\begin{multline}
D_1= \frac{1\!-\!2d\!-\!d^2}{8}+\frac{1}{8(1\!+\!\alpha^2)^3}\left[2(1\!+\!\alpha^2)^2 (d^2\! -\! 2 d\! -\! 5) \right.\\ 
\left.\!+\!4(d\!-\!1) (\alpha\! -\!1)^2(1\!+\!\alpha^2) + 8(\alpha^4\!+\! 6 \alpha^2 \!+\!1)\right]\,,
\end{multline}
\begin{multline}
  D_2=\frac{d(d\!+\!2)}{64}+\frac{1}{32(1\!+\!\alpha^2)^5}
  \left[12 \alpha^3 (1\!+\!\alpha^2) (d\!-\!1)(d\!-\!2)\right.\\
  \left.- 4 \alpha^2 (1\!+\!\alpha^4) (24 \!+\! 4d \!-\! d^2)\right.
  \left.+4 \alpha (1\!+\!\alpha^6) (d\!+\!6) (d\!-\!1) \right.\\
  \left.- (1\!+\!\alpha^8)(26 \!+\!28 d \!+\! 9d^2\right]\,. \label{d1d2}
\end{multline}

\subsection{Stationary state}

In the asymptotic limit, the granular system will reach a stationary state,
that results from the balance between the energy injection by the external
thermostat, and the energy dissipation through inelastic collisions between
the particles. The stationary temperature $T_0$ is thus related both to the
restitution coefficient $\alpha$ and to the amplitude $\xi_0^2$ of the
Gaussian thermostat. Or, to state it differently, in order to ensure a given
value of $T_0$ (for a fixed value of $\alpha$), as resulting from the
stationary form of Eq.~(\ref{nonlinear1}), one has to tune the amplitude
$\xi_0^2$ of the stochastic thermostat to
\begin{multline}
\xi_0^2 =  \frac{n\chi \sigma^{d-1}(1-\alpha^2) S_d }{d \sqrt{\pi}} 
\left( \frac{k_B T_0}{m} \right)^{3/2} \\
\times \left( 1 + \frac{3}{16} a_{20} + \frac{9}{1024} a_{20}^2 \right)\,. 
\label{xi0}
\end{multline} 
Here $a_{20}$ is the stationary value of the coefficient of the first
correction to the Gaussian.  Its expression can be obtained from the
stationary form of Eq~(\ref{nonlinear2}) and it is the solution of the third
order nonlinear equation (see e.g. \cite{BP} for a discussion concerning the
relevance of the corresponding three roots in the case of a force-free
system):
\begin{multline}
(1\!-\!\alpha^2) 
\left( 1\!+\!\frac{3}{16} a_{20} \!+\! \frac{9}{1024} a_{20}^2\right)
\left[1\!+\! a_{20} \frac{(d\!+\!2)(d\!+\!4)}{8} \right] \\
= \sqrt{2}\left( \frac{1\!-\!\alpha^2}{1\!+\!\alpha^2} + D_1\, a_{20}+ 
D_2 \,a_{20}^2\right)\,. 
\label{eqfora20}
\end{multline}

The coefficient $a_{20}$ can be obtained in a closed analytical form through a
Taylor expansion of the above equation. It was however shown in previous
works~\cite{limite,MontaneroSantos} that there are some ambiguities from this
linearization procedure that may affect $a_{20}$. We therefore chose the
linearizing scheme that yields the closest result to the Monte Carlo
simulations of Ref.~\cite{limite}:
\begin{multline}
  a_{20} = - {16(1-\alpha^2)(1+\alpha^2)(1 - \sqrt{2} + \alpha^2)}\\
  \times \left\{16 \sqrt{2} \!+\! 13 \!+\! 4d(3\sqrt{2} \!+\! 1) \!+\! 2 d^2(\sqrt{2}\!-\!1)\right.\\
  \left.\!+\!\alpha^2(\!-\!75 \!+\!44d \!-\! 2 d^2)\!-\! \alpha^4 \left[16 \sqrt{2} \!-\! 3 \!+\! 2 d(d\!+\!6)(\sqrt{2}\!-\!1) \right] \right.\\
  \left.+\alpha^6(-5 + 4 d + 2 d^2)\right\}^{-1}\,.
\end{multline}
Considering instead the expression derived by van Noije and Ernst in
\cite{noije} would not alter the following discussion.  Note that $a_{20}$
becomes zero in the elastic limit $\alpha=1$, when the stationary probability
distribution recovers trivially the Gaussian, equilibrium shape.

The corresponding stationary probability distribution function is therefore
\begin{equation}
f_0(\mathbf{v})=\frac{n}{v_{T_0}^d}\widetilde{f}_0(c)=\frac{n}{v_{T_0}^d}
\phi(c)\left[1+a_{20}S_2(c^2)\right]\,,
\label{scalingfstat}
\end{equation}
where $v_{T_0}=\sqrt{{2k_BT_0}/{m}}$ is the stationary value of the thermal
velocity.

\subsection{Linear stability analysis of the stationary state}

The stability of the aforementioned steady state has not been investigated in
the literature, even if the hydrodynamic-like equations have been derived
recently for the (dilute) system considered here \cite{Garzo}.  A complete 
linear stability analysis (and its eventual comparison with the existing results for 
the homogeneous cooling state~\cite{brey}) is a tedious task, and a separate research subject that we shall not address here further. Instead, we shall consider 
a simplified version of it, in which the homogeneity of the state is not affected by the 
perturbations. This will by no mean influence our general conclusions.

Let us then consider small deviations of the temperature and of the coefficient $a_2$ from
their stationary values,
\begin{equation}
T=T_0(1+\delta \theta)\,,\; a_2=a_{20}+\delta a_2\,,
\label{small}
\end{equation}
with $|\delta \theta| \ll 1\,,\;|\delta a_2| \ll |a_{20}|$.

\begin{figure}[htbp]
\begin{center}
\includegraphics[width=\columnwidth]{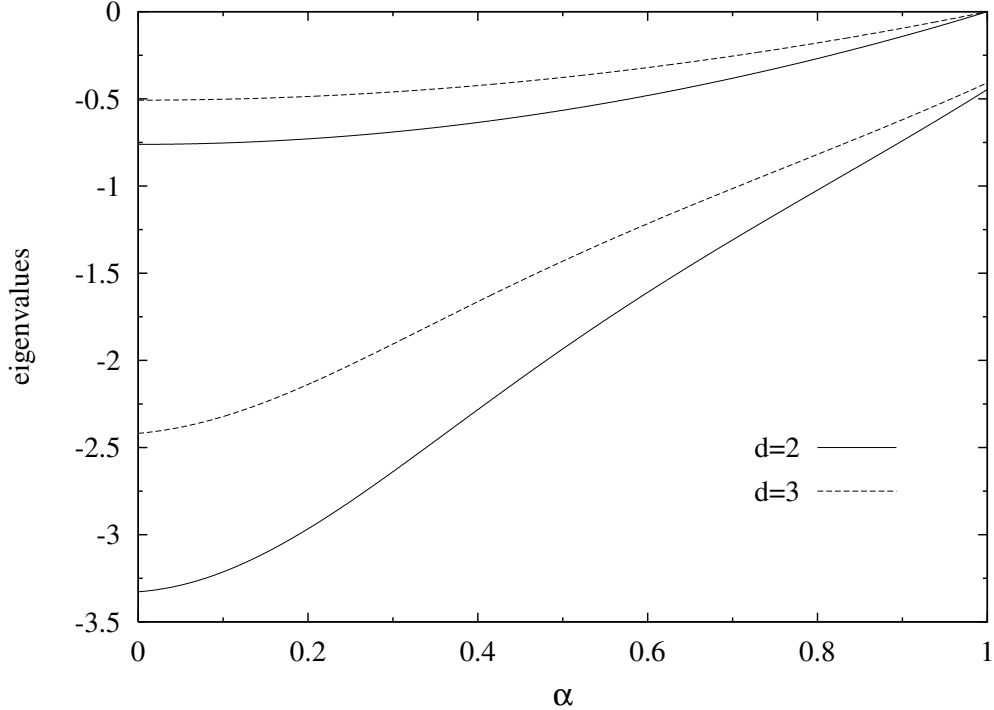}
\end{center}
\caption{The eigenvalues of the linear stability matrix 
for the stationary state as a function of $\alpha$ for $d=2$ and $d=3$. The
eigenvalues are measured in units 
$t_0^{-1}=n\sigma^{d-1}\chi S_d v_{T_0}/\sqrt{2\pi}$.}
\label{eigenvalues}
\end{figure}

 The linearized
evolution equations of these perturbations result from Eqs.~(\ref{nonlinear1})
and (\ref{nonlinear2}),
\begin{multline}
  \frac{d}{dt}(\delta\theta)=\!-\!\frac{m\xi_0^2}{k_BT_0}
  \left[\frac{3}{2}\delta\theta\!+\!\frac{3/16\!+\!(9/512)a_{20}}{1\!+\!(3/16)a_{20}\!+\!
      (9/1024)a_{20}^2}\delta a_2\right]\,,\\
\end{multline}
\begin{multline}
  \frac{d}{dt}(\delta a_2)=-\frac{m\xi_0^2}{k_BT_0} \left\{a_{20}\,\delta
    \theta+\left\{\frac{d+4}{2}+
      \frac{4}{d+2}\right.\right.\\
  \times \left.\left. \left[\left(1+a_{20}\frac{d(d+2)}{8}\right)
        \left(\frac{3}{16}+\frac{9}{512}a_{20}\right)\right.\right.\right.\\
  \left.\left.\left.
        -\frac{\sqrt{2}}{1\!-\!\alpha^2}(D_1\!+\!2D_2a_{20})\right]\left[1\!+\!\frac{3}{16}a_{20}\!+\!
        \frac{9}{1024}a_{20}^2\right]^{-1}\right\}\delta a_2\right\}\,.\\
\end{multline}
In Fig.~\ref{eigenvalues} we have represented the two eigenvalues of the 
corresponding stability matrix as a function of the restitution coefficient 
$\alpha$, for both $d=2$ and $d=3$ cases.

One notices that the two eigenvalues are strictly negative for 
$\alpha < 1$, which indicates the {\em stability} of the stationary 
state with respect to small perturbations in the temperature and in 
the shape of the probability distribution function (in the scaling form).
We emphasize again that spatial homogeneity has been
enforced here. 
As expected, in the elastic limit $\alpha=1$ one of the eigenvalues 
becomes zero (while the other one remains negative) -- which corresponds to 
the temperature becoming a marginal mode,   
and to a relaxation of the distribution function to its equilibrium Gaussian
shape.

\section{Entropy production}

We now turn to the issue of entropy. For our {\em homogeneous}
system, we consider the Shannon information entropy
\begin{equation}
S(t) \equiv  -k_B \int_{\mathbbm{R}^d} \dd \mathbf{v}_1 
f(\mathbf{v}_1,t) \ln \left(\frac{f(\mathbf{v}_1,t)}{eh^d}\right) 
\label{eq5}
\end{equation}
(with Euler's number $e$ and Planck's constant $h$, and $eh^d$ 
the volume of the semiclassical elementary phase-space cell)."  
It is known that in the elastic limit
$\alpha=1$ (and in the absence of an external drive) this reduces to the
appropriate expression of the usual thermodynamic entropy and leads to the
classical ``$H$-theorem".  We now consider the time evolution of $S(t)$ as
governed by the Boltzmann equation~(\ref{eq1}), which reads
\begin{eqnarray}
\frac{d S}{dt}&&=-k_B\int_{\mathbbm{R}^d} \dd \mathbf{v}_1\,
\frac{\partial f(\mathbf{v}_1,t)}{\partial t}\,
\ln\left(\frac{f(\mathbf{v}_1,t)}{eh^d}\right) \nonumber\\
&&=-\chi k_B\int_{\mathbbm{R}^d} 
\dd \mathbf{v}_1 I[f,f] \ln\left(\frac{f(\mathbf{v}_1,t)}{eh^d}\right)
\nonumber\\
&&-\frac{k_B \xi_0^2}{2} \int_{\mathbbm{R}^d} 
\dd \mathbf{v}_1 \frac{\partial^2}{\partial \mathbf{v}_1^2} 
f(\mathbf{v}_1,t) \ln\left(\frac{f(\mathbf{v}_1,t)}{eh^d}\right)\,.
\end{eqnarray}

As mentioned in the Introduction, we wish to introduce a heuristic -- albeit
physically motivated -- entropy production functional that, hopefully,
displays extremal properties in  NESS. 
We shall propose two
approaches. But before proceeding further, we would like to remind the reader
the status of ${d S}/{d t}$ within the framework of phenomenological
thermodynamics as discussed in standard 
textbooks~\cite{degrootmazur,kreuzer,resiboisdeleener}, 
as well as some of its extensions to  stochastic systems~\cite{jiu,bertini,gaspard}.  
Entropy variations are usually split into two
parts:
\begin{equation}
\frac{d S}{d t}={\sigma_{\text{irr}}}+{\sigma_{\text{flux}}}\,,
\end{equation}
where $\sigma_{\text{irr}}\geqslant 0$ is the entropy production arising due
to the dissipative processes that take place inside the system (that is
positively-defined according to the second principle of thermodynamics), while
the entropy flux $\sigma_{\text{flux}}=-\int_VdV\nabla\cdot{\bf J}_S$ accounts
for the external forces driving the system into a nonequilibrium state (the
related contribution is often reduced to boundary terms). The ``art" of
phenomenological thermodynamics precisely bears on ${\bf J}_S$ and on how to
decompose it in terms of the energy, particle, momentum, chemical, {\it etc.},
currents. This is done, usually, on the basis of the local equilibrium
hypothesis. In a similar way, $\sigma_{\text{irr}}$ often appears as a
bilinear form in the fluxes running through the system and the conjugate
affinities. In the near-to-equilibrium regime, the fluxes are usually
proportional to the conjugated affinities, with the Onsager coefficients as
proportionality factors, and one recovers Prigogine's minimum theorem for
$\sigma_{\text{irr}}$ under the hypothesis of time-reversibility of the
underlying microscopic dynamics.

However, in view of the local character of the energy injection mechanism, as
well as of the spatial homogeneity of the system, the situation is completely
different in the case we are considering.  Indeed, unlike the above-mentioned
``conventional'' NESS, there are neither macroscopic, however weak, currents
running across the system, nor the related phenomenological Onsager response
coefficients. Therefore, the separation into ``source" and ``flow" for the
entropy variation is much more tricky.

\noindent {\it First approach.--} A first proposed choice of the ``entropy
production" is
\begin{multline}
\label{sirr}
\sigma_{\text{irr}}=\frac{k_B\chi\sigma^{d-1}}{4}\int\dd \mathbf{v}_1\dd \mathbf{v}_2\dd\widehat{\boldsymbol{\sigma}}\theta(\widehat{\boldsymbol{\sigma}}\cdot{\bf v}_{12})\widehat{\boldsymbol{\sigma}}\cdot{\bf v}_{12}\\
\times \left(f_1^{**}f_2^{**}-f_1 f_2\right) \ln\left(\frac{f_1^{**} f_2^{**}
  }{f_1 f_2}\right)+\frac{\xi_0^2}{2}\int\dd {\bf v}\frac{(\nabla_{\bf v}
  f)^2}{f}\,,
\end{multline}
the form of the first r.h.s. term being simply chosen by analogy with the
elastic-limit case. The second term has been chosen by analogy with
standard diffusion processes. In those processes this term vanishes at
equilibrium because the gradients disappear. Note however that here the
diffusion process happens in the velocity space, and thus the vanishing of
this term at equilibrium is not due to the system becoming homogeneous in
${\bf v}$-space, but because of the energy source strength $\xi_0^2$ being
tuned to $0$. The above $\sigma_{\text{irr}}$ appears to be the sum of two
positive definite terms, and it is therefore also positive definite.
Furthermore, $\sigma_{\text{irr}}$ can only be zero {\em at equilibrium},
namely when both the energy source (the random kicks) and the energy sink (the
dissipative collisions) are tuned to zero. In that respect, it fulfills the
properties expected from standard phenomenological thermodynamics.

On the other hand, the form of the entropy flux $\sigma_{\text{flux}}$ is now
constrained to be
\begin{multline}
  \sigma_{\text{flux}}=\frac{k_B\chi\sigma^{d-1}}{4}\int\dd \mathbf{v}_1\dd \mathbf{v}_2\dd\widehat{\boldsymbol{\sigma}}\theta(\widehat{\boldsymbol{\sigma}}\cdot{\bf v}_{12})\widehat{\boldsymbol{\sigma}}\cdot{\bf v}_{12}\\
  \times f_1 f_2 \ln\left[\frac{(f_1^{**} f_2^{**}) (f_1
      f_2)^{1-\alpha^2}}{(f_1^* f_2^*)^{2-\alpha^2}}\right]\,,
\end{multline}
where we have used the shorthand notations $f_{1,2}=f(\mathbf{v}_{1,2},t)$,
respectively $f_{1,2}^{**}=f(\mathbf{v}_{1,2}^{**},t)$ for the distribution
functions corresponding to the pre-collisional velocities~(\ref{resume4}).
The above functional of $f$ is negative for a large class of trial functions,
and must definitely assume a negative value
$\sigma_{\text{flux}}\sim-\frac{1-\alpha^2}{\ell}T_0^{1/2}$ in the steady
state ($\ell\sim\frac{1}{\chi \sigma^{d-1}}$ is the mean free path). However,
aside from conveying the shrinking of phase space volumes, we must dismiss
$\sigma_{\text{irr/flux}}$ as relevant candidates for extremum entropy
functionals. Indeed, in the spirit of phenomenological thermodynamics, the
splitting of ${d S}/{d t}$ into $\sigma_{\text{irr}}$ and
$\sigma_{\text{flux}}$ is motivated by the desire to isolate the driving
processes (the source and sink referred to above) from the irreversible
processes inside the system. However there is no simple and univoque manner to
do so, and definitely this first choice is not accomplishing this
physically-motivated requirement.  It must be noted that the last term of
eq.(\ref{sirr}) could have also chosen as a part of $\sigma_{\text{flux}}$,
which would then have featured both the source and the sink, at the price of
abandoning its negative definiteness.

\noindent {\it Second approach.--} We now propose an alternative and perhaps 
more pragmatic route, which consists in isolating as the only driving
mechanism the random kicks provided by the thermostat. The inelastic
collisions, viewed above as an energy sink, are now incorporated into a term
describing the system's intrinsic dissipative microscopic dynamics. Along
those lines we henceforth write that
\begin{equation}
\frac{dS}{dt}=\sigma_{syst}+\sigma_{ext}\,.
\end{equation}
The first contribution $\sigma_{syst}$
corresponds to the entropy production 
inside the system,
i.e., it comes from the changes of the particles velocities during the 
binary inelastic collisions, 
\begin{multline}
\sigma_{syst}=\frac{k_B\chi\sigma^{d-1}}{2} 
\int_{\mathbbm{R}^d} \dd \mathbf{v}_1 \int_{\mathbbm{R}^d} 
\dd \mathbf{v}_2 \int \dd \widehat{\boldsymbol{\sigma}} 
 \theta(\widehat{\boldsymbol{\sigma}} \cdot \mathbf{v}_{12}) \\
\times (\widehat{\boldsymbol{\sigma}} \cdot \mathbf{v}_{12}) f_1 f_2 \, \ln \left(\frac{f_1 f_2}{f_1^* f_2^*}\right)\,,
\end{multline}
where we have used the shorthand notation
$f_{1,2}^*=f(\mathbf{v}_{1,2}^*,t)$ for the distribution functions 
corresponding to the post-collisional
velocities~(\ref{resume42}). Of course, in the limit of elastic collisions
$\alpha=1$ the expression of $\sigma_{syst}$ reduces to the usual
positive-definite expression of the hard-disk gas that enters the 
$H$-theorem. However, in general $\sigma_{syst}$ does not have a definite sign.
One can imagine the entropy production inside the system as resulting from two 
antagonist (although actually undissociated) mechanisms, namely a generic disordering effect
of any particle collisions (e.g., that is also present for elastic hard spheres) 
in $d\geqslant 2$,
and an ordering effect due to the inelastic character of the collisions
(i.e., to the reduction of the translational agitation of the particles). 
Depending on the actual shape of the distribution function, one of these two mechanisms 
may prevail on the other, thus determining the sign of the instantaneous value
of $\sigma_{syst}$.

The second contribution $\sigma_{ext}$ is determined by the 
effect of the thermostat on the distribution function of 
the particles of the system. 
It corresponds to an energy injection into the system, and to a 
disordering effect of the particles velocities (through 
``random kicking"), and therefore, as expected, is always a 
{\em positively-defined}
quantity,
\begin{equation}
\sigma_{ext}=\frac{k_B \xi_0^2}{2} \int_{\mathbbm{R}^d}
\dd \mathbf{v}\, \frac{1}{f(\mathbf{v},t)} 
\left[ \boldsymbol{\nabla}_{\mathbf{v}} f(\mathbf{v},t) \right]^2\,.
\label{sd}
\end{equation}

Introducing the dimensionless quantities 
\begin{equation}
\widetilde{\sigma}_{syst,ext}=\frac{2\sigma_{syst,ext}}{\chi \sigma^{d-1}
v_{T_0} n^2}\,,
\end{equation}
one obtains the expressions for 
the dimensionless time-dependent entropy production sources:
\begin{multline}
\widetilde{\sigma}_{syst} = \left[ \frac{T(t)}{T_0} 
\right]^{1/2} \int_{\mathbbm{R}^d} \dd \mathbf{c}_1 
\int_{\mathbbm{R}^d} \dd \mathbf{c}_2 \int \dd   
\widehat{\boldsymbol{\sigma}} \, 
\theta(\widehat{\boldsymbol{\sigma}}\cdot \mathbf{c}_{12}) \\
\times
(\widehat{\boldsymbol{\sigma}} \cdot \mathbf{c}_{12})\,  
\widetilde{f}(\mathbf{c}_1,t)\widetilde{f}(\mathbf{c}_2,t) \,
\ln \left[ \frac{\widetilde{f}(\mathbf{c}_1,t)\widetilde{f}(\mathbf{c}_2,t)}
{\widetilde{f}(\mathbf{c}_1^*,t)\widetilde{f}(\mathbf{c}_2^*,t)}
\right]\,,
\label{s12}
\end{multline}
respectively
\begin{multline}
\widetilde{\sigma}_{ext} = 
\left[\frac{T_0}{T(t)}\right] \frac{(1\!-\!\alpha^2)S_d}{2d \sqrt{2 \pi}} 
\left( 1 \!+\! \frac{3}{16} a_2(t) \!+\! \frac{9}{1024} a_2^2(t)\right) \\
\times
\int_{\mathbbm{R}^d} \dd \mathbf{c}
\, \frac{1}{\widetilde{f}(\mathbf{c},t)} 
 \left[ \boldsymbol{\nabla}_{\mathbf{c}} \widetilde{f}(\mathbf{c},t) 
 \right]^2\,.
\label{sd2}
\end{multline}

\begin{figure}[htbp]
\begin{center}
\includegraphics[width=\columnwidth]{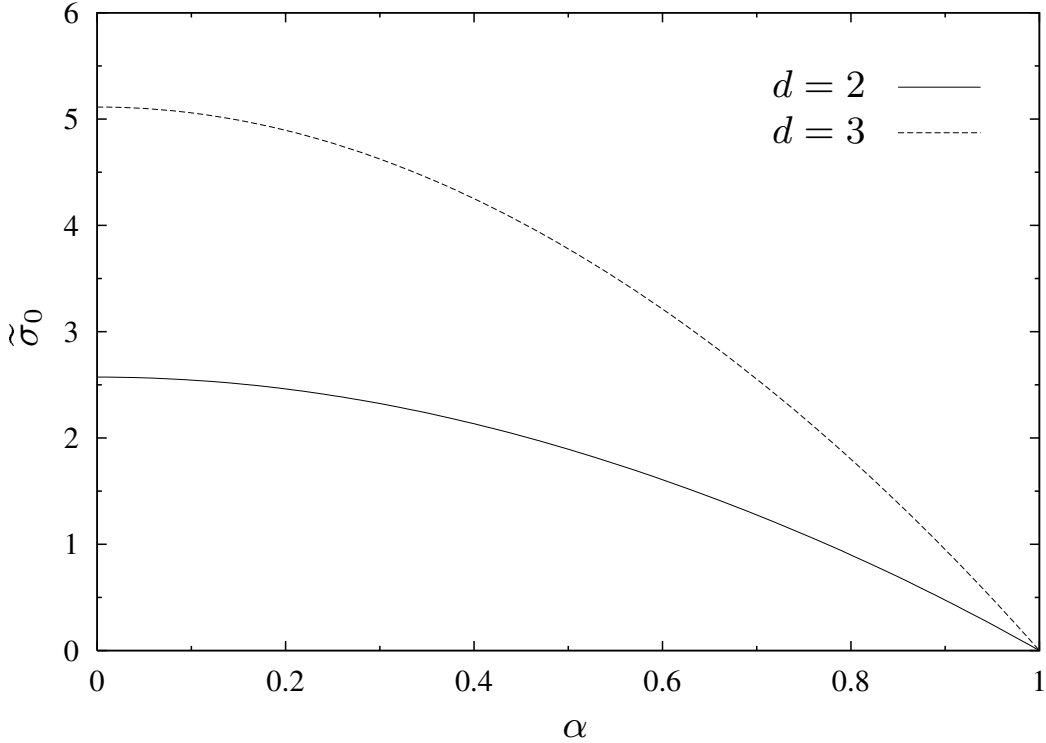}
\end{center}
\caption{The dimensionless stationary entropy production 
$\widetilde{\sigma}_0$ as a function of $\alpha$ in $d=2$ and $d=3$.}
\label{sigma0}
\end{figure}

In the stationary regime at temperature $T_0$ one has, obviously,
$\widetilde{\sigma}_{syst}=-\widetilde{\sigma}_{ext}
\equiv -\widetilde{\sigma}_0$. The quantity $\widetilde{\sigma}_0$ is positive
and decaying monotonously with $\alpha$, as illustrated in  
Fig.~\ref{sigma0}. 
Note that $\widetilde{\sigma}_0$  is nonzero as
long as the collisions are inelastic, i.e., as long as the stationary
probability distribution is non-Gaussian. Note also the negativity of 
$\widetilde{\sigma}_{syst}$
in the stationary state -- the ordering effect due to 
the inelastic character of the 
collisions prevails on the generic disordering 
effect of the collisions.

Let us now address the question whether the entropy production 
(as a whole, or one of its parts $\widetilde{\sigma}_{syst}$
or $\widetilde{\sigma}_{ext}$) can play the role of some kind 
of ``nonequilibrium potential" for the system, i.e., whether or not
it can account for the linear stability of the stationary state 
of the system. 
The particular case of the quasi-elastic limit 
$\varepsilon \equiv 1-\alpha \ll 1$ 
is especially interesting, given that the stationary state is close to equilibrium.
One might then  expect {\it a priori}
that a ``minimum entropy production theorem" 
(in the spirit of the ``extended Prigogine theory''~\cite{jiu}) 
might be valid in this case.

Consider thus small perturbations of the temperature and of the 
coefficient $a_2$ around their stationary values, as 
in Eq.~(\ref{small}). A Taylor development of the entropy production 
terms $\widetilde{\sigma}_{syst}$ and $\widetilde{\sigma}_{ext}$
leads to nonzero linear contributions in the perturbations 
$\delta \theta$ and $\delta a_2$,
\begin{multline}
\widetilde{\sigma}_{syst}-(-\widetilde{\sigma}_0)=-\delta \theta 
\,\left(\frac{\widetilde{\sigma}_0}{2}\right)+ \delta a_2 \,
\int_{\mathbbm{R}^d} \dd \mathbf{c}_1 
\int_{\mathbbm{R}^d} \dd \mathbf{c}_2 \int \dd   
\widehat{\boldsymbol{\sigma}} \\
\times \theta(\widehat{\boldsymbol{\sigma}}\cdot \mathbf{c}_{12}) 
(\widehat{\boldsymbol{\sigma}} \cdot \mathbf{c}_{12})\, 
\widetilde{f}_0(c_1) \widetilde{f}_0(c_2)\, 
\left\{\left[ \frac{S_2(c_1^2)}{1+a_{20}S_2(c_1^2)}+\right.\right.\\
\times \left.\left.
\frac{S_2(c_2^2)}{1+a_{20}S_2(c_2^2)}
-\frac{S_2(c_1^{*2})}{1+a_{20}S_2(c_1^{*2})}-\frac{S_2(c_2^{*2})}
{1+a_{20}S_2(c_2^{*2})}\right]\right. \\
+\left.\left[ \frac{S_2(c_1^2)}{1+a_{20}S_2(c_1^2)}+
\frac{S_2(c_2^2)}{1+a_{20}S_2(c_2^2)}\right]\,
\ln \left[ \frac{\widetilde{f}_0(\mathbf{c}_1)
\widetilde{f}_0(\mathbf{c}_2)}
{\widetilde{f}_0(\mathbf{c}_1^*)\widetilde{f}_0(\mathbf{c}_2^*)} 
\right]\right\} \\
+{ O}(\delta \theta^2, \delta a_2^2)\,,
\end{multline}
respectively
\begin{multline}
\widetilde{\sigma}_{ext}-(\widetilde{\sigma}_0)=-\delta \theta 
({\widetilde{\sigma}_0})+ \delta a_2
\Bigg\{\displaystyle\frac{3/16\!+\!(9/512)a_{20}}{1\!+\!(3/16) a_{20}\!+\!(9/1024)
a_{20}^2}\,\widetilde{\sigma}_0\Bigg.\\
\Bigg.+\frac{(1\!-\!\alpha^2) S_d}{2d\sqrt{2\pi}}
\left(1\!+\!\frac{3}{16}a_{20}\right.\Bigg.
\left.\!+\!\frac{9}{1024}a_{20}^2\right)
\Bigg.\\
\times \Bigg.
\int_{\mathbbm{R}^d} \dd \mathbf{c} 
\left[\frac{2\left(\boldsymbol{\nabla}_{\mathbf{c}} 
\widetilde{f}_0(\mathbf{c})\right)
\cdot \left[\boldsymbol{\nabla}_{\mathbf{c}}
\left(e^{-c^2}S_2(c^2)\right)\right]}{\pi^{d/2}\widetilde{f}_0(\mathbf{c})}\right.\Bigg.\\
\left.\Bigg.
-\frac{\left(\boldsymbol{\nabla}_{\mathbf{c}} 
\widetilde{f}_0(\mathbf{c})\right)^2\,\left(e^{-c^2}S_2(c^2)\right)}
{\pi^{d/2}\widetilde{f}_0^{\,2}(\mathbf{c})}
\right]\Bigg\}
+{O}(\delta \theta^2, \delta a_2^2)\,.
\end{multline}
The total entropy production 
$\widetilde{\sigma}_{syst}+\widetilde{\sigma}_{ext}$ 
also contains linear terms in the perturbations 
$\delta \theta$ and $\delta a_2$. 

The same holds true even in the quasielastic limit 
$\varepsilon \equiv 1-\alpha \ll 1$,
when one can evaluate explicitly to 
${O}(\varepsilon ^2)$ the expression of the coefficients 
of the perturbations. More precisely,
\begin{multline}
\widetilde{\sigma}_{ext}-(-\widetilde{\sigma}_0)=-\delta \theta 
\,\left(\frac{\widetilde{\sigma}_0}{2}\right) - \delta a_2 \,
\frac{\sqrt{2}\,\pi^{\frac{d-1}{2}}}{\Gamma(d/2)}\\
\times \left[2(d-1)\,a_{20}+\frac{4d+5}{8}\, \varepsilon\,+ 
{O}(\varepsilon ^2)\right]\,,\\
\widetilde{\sigma}_{syst}-\widetilde{\sigma}_0=-\delta \theta 
\,\left({\widetilde{\sigma}_0}\right) +\delta a_2\,
\left[\frac{3}{16}\, \widetilde{\sigma}_0 \,+\,
{ O}(\varepsilon ^2)\right]\,,
\end{multline}
where the stationary values are
\begin{equation}
\widetilde{\sigma}_0 = \frac{2\sqrt{2}\pi^{ \frac{d-1}{2} }}
{\Gamma(d/2)}\,\varepsilon \,+\,{ O}(\varepsilon ^2)\,,
\end{equation}
and
\begin{equation}
a_{20} =-\frac{\sqrt{2}(\sqrt{2}-1)}{d-1}\,\varepsilon\,+\,
{ O}(\varepsilon ^2)\,.
\end{equation}

The meaning of this result is that the entropy production as defined above
cannot be used for a variational description of the relaxation of the system
towards the stationary state, not even in the quasi-elastic limit. 

One may argue that the choice of the definition of the entropy 
production inside the system might be inappropriate, 
since it refers only to the translational degrees of freedom,
and it does not take into 
account properly the internal degrees
of freedom of the particles -- that are, in fact,
responsible for the inelastic character of
the collisions. 
The description of the inelasticity through a
constant restitution coefficient $\alpha$ 
might thus be incompatible
with a thermodynamic
description of the system in terms of entropy production.
We note that it is known that such a model, although being a useful 
approximation which captures important physical effects,
is in fact incompatible with basic mechanical laws
(see e.g. chapter 3 of Ref. \cite{BP}). 

Let us now discuss briefly another issue that draw attention 
recently, see Refs.~\cite{mackey05,piotr05}, 
namely that of the {\em Kullback relative entropy}, defined as
\begin{eqnarray}
S_R(t)&=&-k_B\int_{\mathbbm{R}^d} \dd \mathbf{v} 
f(\mathbf{v},t) \ln \left(\frac{f(\mathbf{v},t)}{f_0(\mathbf{v})}\right) \nonumber\\
&=&-\frac{k_Bn}{v_T^d}\int_{\mathbbm{R}^d} \dd \mathbf{c}\; \phi(c)\left[1+a_2S_2(c^2)\right]\,\mbox{ln}\left(\frac{v_{T_0}^d}{v_T^d}\;\frac{1+a_2S_2(c^2)}{1+a_{20}S_2(c^2)}\right)\,.
\end{eqnarray}
$S_R(t)$ is a measure of the ``distance" between the actual pdf $f(\mathbf{v},t)$
and its stationary profile $f_0(\mathbf{v})$, and, of course, is equal to zero at the stationary state. Following Ref.~\cite{piotr05}, one can parametrize $S_R(t)$
through the  two sets of parameters, $\{\gamma_1=a_2(t), \gamma_2=T(t)\}$ for the nonstationary state, respectively $\{\gamma_{10}=a_{20}, \gamma_{20}=T_0\}$ for the sationary pdf. Considering as above (sec.~ 3.3) 
small deviations of the temperature and of the coefficient $a_2$ 
(that result in a small deviation $\delta f (\mathbf{v},t)$ of the pdf ) from their stationary values, one finds:
\begin{eqnarray}
\delta S_R&\approx&-\frac{k_B}{2}\int_{\mathbbm{R}^d} \dd \mathbf{v} \frac{1}{f_0(\mathbf{v})}
[\delta f (\mathbf{v},t)]^2=-\frac{1}{2} \sum_{i,j=1,2} { F}_{ij} \delta \gamma_1 \delta \gamma_2 \leqslant 0\,,
\end{eqnarray}
where $\delta \gamma_1= \delta a_2$, $\delta \gamma_2= T_0 \delta \theta$,  
and ${ F}_{ij}$ is the positively-defined Fisher information matrix~\cite{fisher},
\begin{equation}
{ F}_{ij}=k_B \int_{\mathbbm{R}^d} \dd \mathbf{v} f_0(\mathbf{v})\;
\left(\frac{\partial \mbox{ln} f_0(\mathbf{v})}{\partial \gamma_{i0}}\right)\;
\left(\frac{\partial \mbox{ln} f_0(\mathbf{v})}{\partial \gamma_{j0}}\right)\,.
\end{equation}
It looks therefore as if this relative entropy has the required property of extremum at the steady-state (and monotonous exponential asymptotic relaxation towards it). 
This property has already been demonstrated for other types of nonequilibrium stochastic systems (e.g., in Ref.~\cite{mackey05}, the one-dimensional Ornstein-Uhlenbeck and Rayleigh processes, noise-perturbed harmonic oscillator, dichotomous noise). The question arises about its relationship with the thermodynamic entropy production; in Ref.~\cite{piotr05} it was shown that in the case of the usual Smolukhowski diffusion 
the Kullback entropy time-variation rate coincides with the Shannon entropy 
production rate. However, some further case-study (in particular, on systems described by kinetic Boltzmann-like equations) are necessary 
before generalizing this important conclusion to other nonequilibrium 
situations. In particular,
although very appealing, the Kullback entropy does not reduce to the usual 
$H$-functional in the limit of an elastic gas of particles relaxing to 
equilibrium. Besides that, computing Kullback entropy requires the knowledge
of the steady-state pdf, while the expected approach would be to define 
a proper Lyapunov functional of the system from which to {\em deduce} the
stationary state.

\section{Conclusions}

We illustrated on the well-known 
model of a randomly driven granular gas with constant restitution coefficient the difficulties that one encounters when trying to construct a 
variational principle for  NESS based on an 
``entropy production''. Two approaches were proposed for the interpretation of the
entropy balance equation in terms of ``sources" and ``flows", but none of them lead to 
the formulation of such a principle. The main reason for this failure seems to  be
the intrinsic irreversible microscopic dynamics of the granular gas.
Modeling the internal degrees of freedom of the grains (that are responsible for the 
inelasticity of the collisions) through a constant restitution coefficient is thus not appropriate for a thermodynamic description. 
This shows thus a major limitation of this model.
A further step in the rather involved question of the refinement of the description
would be the use of {\em random restitution coefficients}
(as done, e.g., in~\cite{zippelius}). These are meant to describe the possible flow of energy (at the collision) both towards and {\em from} the internal degrees of freedom to 
the translational degrees of freedom. Such a model, however, cannot be treated analytically, and no simple analytic conclusions can be therefore drawn on the fate of the corresponding $H$-functional. Numerical results are left for further studies.

Moreover, the problem of the Kullback relative entropy, its monotonous 
relaxation to the steady-state, and its relationship with the thermodynamic entropy production of a nonequilibrium system is a very promising direction for further studies.

We thank Ph. Martin and J. Piasecki for suggestions and
discussions. I.B., F.C., and M.D. acknowledge partial support from the Swiss
National Science Foundation.

\end{document}